\documentclass{article}

\usepackage{PRIMEarxiv}

\usepackage{amsmath}
\usepackage[utf8]{inputenc} % allow utf-8 input
\usepackage[T1]{fontenc}    % use 8-bit T1 fonts
\usepackage{hyperref}       % hyperlinks
\usepackage{url}            % simple URL typesetting
\usepackage{booktabs}       % professional-quality tables
\usepackage{threeparttable} % 
\usepackage{float}
\usepackage{amsfonts}       % blackboard math symbols
\usepackage{nicefrac}       % compact symbols for 1/2, etc.
\usepackage{microtype}      % microtypography
\usepackage{lipsum}
\usepackage{fancyhdr}       % header
\usepackage{graphicx}       % graphics
\usepackage{listings}       % Enhanced syntax highlighting and code appearance
\graphicspath{{media/}}     % organize your images and other figures under media/ folder

\title{GeneralizIT: A Python Solution for Generalizability Theory Computations
\thanks{\textit{\underline{Citation}}: 
\textbf{}} 
}

\author{
  Tyler J. Smith$^{1,2,3}$
  Theresa Kline$^{4}$, 
  Adrienne Kline$^{1,2,3,*}$ \\
  $^1$Center for Artificial Intelligence, Northwestern Medicine, Chicago \\
  $^2$Divison of Cardiac Surgery, Northwestern University, Chicago \\
  $^2$Dept. of Electrical and Computer Engineering, Northwestern University, Chicago \\
  $^4$University of Calgary, Calgary \\
  $*$Corresponding author\\
}

\begin{document}
\maketitle

\begin{abstract}

\texttt{GeneralizIT} is a Python package designed to streamline the application of Generalizability Theory (G-Theory) in research and practice.
G-Theory extends classical test theory by estimating multiple sources of error variance, providing a more flexible and detailed approach to reliability assessment. Despite its advantages, G-Theory's complexity can present a significant barrier to researchers. \texttt{GeneralizIT} addresses this challenge by offering an intuitive, user-friendly mechanism to calculate variance components, generalizability coefficients ($E\rho^2$) and dependability ($\Phi$) and to perform decision (D) studies. D-Studies allow users to make decisions about potential study designs and target improvements in the reliability of certain facets. The package supports both fully crossed and nested designs, enabling users to perform in-depth reliability analysis with minimal coding effort. With built-in visualization tools and detailed reporting functions, \texttt{GeneralizIT} empowers researchers across disciplines, such as education, psychology, healthcare, and the social sciences, to harness the power of G-Theory for robust evidence-based insights. Whether applied to small or large datasets, \texttt{GeneralizIT} offers an accessible and computationally efficient solution to improve measurement reliability in complex data environments.

\end{abstract}

% keywords can be removed

\section{Statement of Need}
Generalizability Theory (G-Theory) offers a powerful extension to Classical Test Theory by allowing the estimation of multiple sources of error variance, providing a more nuanced and comprehensive approach to assessing measurement reliability \cite{brennan:2010}. However, its inherent complexity often limits its accessibility, particularly for researchers who lack advanced statistical training or coding expertise \cite{teker:2015}. This presents a significant barrier to its widespread adoption. 

Currently, no Python-based package exists for conducting Generalizability Theory (G-Theory) analyses. Current implementations require proprietary or specialized statistical software such as SAS, SPSS, EduG, G-String-MV \cite{bloch:2012, bloch:2024}, or the R package \texttt{gtheory} \cite{moore:2016}. Table~\ref{tab:gtheory-software} provides an updated summary of the available software for G-Theory analyses \cite{briesch:2014}. These tools require steep learning curves that may not be accessible to the growing number of Python-based researchers, leaving many relying on less robust methods for reliability analysis.

\begin{table}[H]
\centering
\begin{threeparttable}
\caption{Summary of available software for G-Theory analyses}
\label{tab:gtheory-software}
\begin{tabular}{@{}lcccp{.1\textwidth}cc@{}}
\toprule
\textbf{Software} & \textbf{Availability} & \textbf{Input} & \textbf{Missing Data} & \textbf{Unbalanced Designs}\textsuperscript{a} & \textbf{D Studies} \\ \midrule
\texttt{G\_String\_MV}\textsuperscript{b} & Free from \href{https://github.com/G-String-Legacy/GS_MV/releases}{GitHub} & Raw data & No\textsuperscript{c} & Yes & Yes \\
SAS VARCOMP & Included in SAS & Raw data & Yes\textsuperscript{d} & Yes & No \\
SPSS VARCOMP & Included in SPSS & Raw data & Yes\textsuperscript{e} & Yes & No \\
EduG & Free from \href{http://www.irdp.ch/edumetrie/englishprogram.htm}{IRDP} & Raw data or sums of squares & No & No & Yes \\
\texttt{gtheory} & R package & Raw data & No & Yes & Yes \\
\texttt{generalizIT} & Python package & Raw data & No & No & Yes \\ \bottomrule
\end{tabular}
\begin{tablenotes}
\item[a] Unbalanced design indicates that each level of the nesting facet includes a different number of levels of the nested facet (e.g., a variable number of observations within each day). 
\item[b] Built upon urGENOVA.
\item[c] Mean substitutions if missing data detected prior to analysis
\item[d] Listwise deletion for missing dependent variable data.
\item[e] Listwise deletion if any dependent variable value is missing.
\end{tablenotes}
\end{threeparttable}
\end{table}

The \texttt{GeneralizIT} Python package addresses this critical need by offering a user-friendly platform that simplifies the application of G-Theory in research and practice. By automating the calculation of variance components, generalizability coefficients, and dependability indices, \texttt{GeneralizIT} makes this sophisticated theory accessible to a broad range of disciplines, including education, psychology, healthcare, and the social sciences. The package supports crossed experimental designs for any number of facets and nested designs up to two facets of differentiation, enabling users to conduct detailed reliability analyses without extensive coding knowledge. Additionally, its built-in visualization and reporting tools provide clear, interpretable outputs, further enhancing its utility.

In a research landscape where measurement reliability is paramount for producing valid, evidence-based conclusions, GeneralizIT fills an urgent need for a computationally efficient and accessible solution. It democratizes the use of G-Theory, allowing researchers to obtain more reliable insights even from complex, small, or large datasets.

\section{Usage}

\texttt{GeneralizIT} is designed to be user-friendly and accessible to researchers across disciplines. The package provides a simple interface for conducting Generalizability Theory analyses, including calculating variance components, generalizability and dependability coefficients, conducting D-Studies, and generating confidence intervals. The following sections outline the key functionalities of \texttt{GeneralizIT}:
\begin{itemize}
    \item \textbf{Installation}
    \item \textbf{Input Data}
    \item \textbf{Calculating Variance Components}
    \item \textbf{Generalizability and Dependability Coefficients}
    \item \textbf{D-Studies}
    \item \textbf{Confidence Intervals}
    \item \textbf{Summary Statistics}
\end{itemize}

\subsection{Installation}
Install the \texttt{GeneralizIT} package using pip:

   \texttt{\$ pip install generalizit}

\subsection{Input Data}

\subsubsection{Importing the Package}
\texttt{from generalizit import GeneralizIT}

\subsubsection{Preparing the Data}
Data should be input as a flat Pandas DataFrame with columns representing the unique facets and response. 
For example, a csv dataset of a fully crossed design with facets \texttt{person}, \texttt{item}, \texttt{rater}, and a response column can read in as follows:

\begin{verbatim}
read in the data
data = pd.read_csv("data.csv")
print(data.head(8))
print(data.tail(8))
\end{verbatim}

\begin{lstlisting}
| Person | item | rater | Response |
|--------|------|-------|----------|
|      1 |    1 |     1 |        2 |
|      1 |    2 |     1 |        6 |
|      1 |    3 |     1 |        7 |
|      1 |    4 |     1 |        5 |
|      1 |    1 |     2 |        2 |
|      1 |    2 |     2 |        5 |
|      1 |    3 |     2 |        5 |
|      1 |    4 |     2 |        5 |
% ...
|     10 |    1 |     1 |        6 |
|     10 |    2 |     1 |        8 |
|     10 |    3 |     1 |        7 |
|     10 |    4 |     1 |        6 |
|     10 |    1 |     2 |        6 |
|     10 |    2 |     2 |        8 |
|     10 |    3 |     2 |        8 |
|     10 |    4 |     2 |        6 |
\end{lstlisting}
Conversely, if the design was nested such as $person \times (rater:item)$, raters are nested under item and should be identified uniquely either by delineation \texttt{item1\_rater1} or unique numbering as below:
\begin{lstlisting}
| Person | item | rater | Response |
|--------|------|-------|----------|
|      1 |    1 |     1 |        2 |
|      1 |    1 |     2 |        6 |
|      1 |    1 |     3 |        7 |
|      1 |    1 |     4 |        5 |
|      1 |    2 |     5 |        2 |
|      1 |    2 |     6 |        5 |
|      1 |    2 |     7 |        5 |
|      1 |    2 |     8 |        5 |
|      1 |    3 |     9 |        6 |
|      1 |    3 |    10 |        8 |
|      1 |    3 |    11 |        7 |
|      1 |    3 |    12 |        6 |
|      2 |    1 |     1 |        6 |
|      2 |    1 |     2 |        8 |
|      2 |    1 |     3 |        8 |
|      2 |    1 |     4 |        6 |
% ...
\end{lstlisting}

\subsubsection{Initialize the GeneralizIT Class}
\begin{verbatim}
# Create a Generalizability object
GT = GeneralizIT(data=data, design_str="person x rater x item", response="Response")
\end{verbatim}

\textbf{Note:}
\begin{itemize}
  \item The \texttt{data} parameter should be a Pandas DataFrame containing the data as described above. The data must be balanced, fully crossed, and must not contain missing values.
  \item The \texttt{design\_str} parameter should be a string representing the design of the study. For designs with crossed facets, the facets should be separated by \texttt{" x "}, as shown in the example above. For nested designs, the facets should be separated by \texttt{:}. If there is a mixed design, it is important to include \texttt{()} around the appropriate facets. For example, a mixed design with a nested facet could be written as \texttt{Person x Item:Rater}. However, the interpretations \texttt{Person x (Item:Rater)} or \texttt{(Person x Item):Rater} result in different designs and different calculations, thus it is important to be explicit in the design string.
  \item The \texttt{response} parameter should be a string representing the column name of the response variable in the data.
\end{itemize}

\subsection{Calculating Variance Components}
Variance components including sum of squares, mean squares and $\sigma^2$ for each facet and combination of facet interactions can be calculated using the \texttt{calculate\_anova()} method.
\begin{verbatim}
# Calculate variance components
GT.calculate_anova() 
\end{verbatim}

\subsection{Generalizability and Dependability Coefficients}
Calculate generalizability, $E\rho^2$, and dependability, $\Phi$, coefficients using the \texttt{g\_coeffs()} method. 
\begin{verbatim}
# Differentiation table for generalizability and dependability coefficients
GT.g_coeffs()  
\end{verbatim}

\subsection{D-Studies}
Conduct D-Studies to estimate the reliability of measurement instruments under differing levels using the \texttt{calculate\_d\_study()} method. 
\begin{verbatim}
#Perform a D-Study
GT.calculate_d_study(levels = {'person':[10], 'i': [4, 8], 'o': [1,2]})  
\end{verbatim}
\textbf{Note:}
\begin{itemize}
    \item \texttt{levels} is a dictionary with keys as the facets and values as a list of potential levels for each facet. 
\end{itemize}

\subsection{Confidence Intervals}
Calculate confidence intervals for the expected values of each facet using the \texttt{calculate\_confidence\_intervals()} method. 
\begin{verbatim}
# Get the confidence intervals for each potential object of measurement's mean scores
GT.calculate_confidence_intervals(alpha=0.05)
\end{verbatim}

\textbf{Note:}
\begin{itemize}
    \item $\alpha$ is the significance level for the confidence intervals. Default is $0.05$.
\end{itemize}

\subsection{Summary Statistics}
Print summary statistics including ANOVA table (\texttt{df}, \texttt{T}, \texttt{SS}, \texttt{MS}, $\sigma^2$), G coefficients table, D-Study tables, and confidence intervals using the following methods:
\begin{verbatim}
# Summary Statistics
GT.anova_summary()  # Print ANOVA table
GT.g_coeff_summary()  # Print differentiation table
GT.d_study_summary()  # Print D-Study results
GT.confidence_intervals_summary()  # Print confidence intervals
\end{verbatim}

\section{Methods}
The package relies on equations from \cite{brennan:2001} and \cite{cardinet:1976} to calculate variance components, generalizability and dependability coefficients, confidence intervals, and D-Studies. The following sections provide an overview of the key equations and methods used in GeneralizIT.

\subsection{Fully Crossed Designs}
\subsubsection{Calculating Sum of Squares and Mean Squares for Variance Components}
\[
\text{First, Calculate T Values:}
\]
\[
T(\alpha) = \pi(\alpha^*) \sum \left( \bar{\alpha} \right)^2
\]
\[
\text{where } \pi(\alpha^*) \text{ is the product of levels of all facets except } \alpha.
\]

\[
\text{Then:}
\]
\[
SS(\alpha) = T(\alpha) + (-1)^1 \left( \sum T(\beta) \text{ for } \beta \in \alpha \right) 
+ (-1)^2 \left( \sum T(\gamma) \text{ for } \gamma \in \alpha \right) - \ldots 
+ (-1)^n T(U)
\]
\[
\text{where } n \text{ is the number of facets in } \alpha,
\]
\[
\beta \text{ is a subset of interactions of length } n - 1 \text{ facets in } \alpha,
\]
\[
\gamma \text{ is a subset of interactions of length } n - 2 \text{ facets in } \alpha,
\]

\[
\text{Finally Mean Squares:}
\]
\[
MS(\alpha) = \frac{SS(\alpha)}{Df(\alpha)}
\]

\[
\text{For example:}
\]
\[
SS(AB) = T(AB) - \big(T(A) + T(B)\big) + T(U).
\]
\[
MS(AB) = \frac{SS(AB)}{Df(AB)}
\]

\subsubsection{Calculating Variance Components}

\[
\sigma^2(\alpha) = \frac{1}{\pi(\alpha^*)} \left[ \text{linear combination of mean squares (MS)} \right]
\]
\[
\text{where } \pi(\alpha^*) \text{ is the product of levels of all facets except } \alpha.
\]
\[
\text{The linear combination of mean squares is determined by the following algorithm:}
\]
\[
\begin{aligned}
&\text{Identify all components that consist of the } t \text{ indices in } \alpha \text{and exactly one additional index;} \\
&\text{ and call the set of "additional" indices } A.
\end{aligned}
\]

\[
\begin{aligned}
&\text{Step 0: } MS(\alpha), \\
&\text{Step 1: } - \text{mean squares for all components that consist of } t \text{ indices in } \alpha \\
&\quad \text{and exactly one of the indices in } A, \\
&\text{Step 2: } + \text{mean squares for all components that consist of } t \text{ indices in } \alpha \\
&\quad \text{and any two of the indices in } A, \\
&\vdots \\
&\text{Step } n: + (-1)^n \bigg[\text{mean squares for all components that consist of } t \text{ indices in } \alpha \\
&\quad \text{and } n \text{ of the indices in } A \bigg].
\end{aligned}
\]

\[
\text{For example:}
\]
\[
\sigma^2(a) = \frac{MS(a) - MS(ab) - MS(ac) + MS(abc)}{b \cdot c}.
\]

\[
\text{Where } b \text{ and } c \text{ are the levels for the respective facets}
\]

\subsection{Nested Designs}
\texttt{GeneralizIT} allows for designs with nesting for designs with 1 object of measurement and 1 or 2 facets of differentiation. For these designs, the user provided string is compared to tables in appendix A and B of \cite{brennan:2001} to properly calculate variance components.

\subsection{Generalizability and Dependability Coefficients}
\label{sec:G-coeff}

\[
E\rho^2 = \frac{\sigma^2(\tau)}{\sigma^2(\tau) + \sigma^2(\delta)}
\]

\[
\sigma^2(\tau) = \sigma^2(\alpha) + \frac{\sigma^2(\alpha_{\text{fixed}})}{n_{\text{fixed}}}
\]
\[
\text{where } \alpha_{\text{fixed}} \text{ represents all interaction variances between } \alpha \text{ and ONLY fixed factors.}
\]

\[
\sigma^2(\delta) = \frac{\sigma^2(\alpha_{\text{random}})}{n_{\text{random}}}
\]
\[
\text{where } \alpha_{\text{random}} \text{ represents all interaction variances between } \alpha \text{ and ONLY random factors.}
\]

\[
\Phi = \frac{\sigma^2(\tau)}{\sigma^2(\tau) + \sigma^2(\Delta)}
\]
\[
\text{where } \sigma^2(\Delta) \text{ is the sum of all variances } \sigma^2(\bar{\alpha}) \text{ except the variance of } \sigma^2(\tau).
\]

\subsection{D-Study}
To conduct a D-Study, $\sigma^2(\alpha)$ calculated from the G-Study is held constant for each facet and interactions, $\alpha$, while the levels for each facet is varied according to user input. The formulas from \ref{sec:G-coeff} are then repeated for each potential combination to return G-Coefficients for each potential G Study Design. 

\subsection{Confidence Intervals}
Variation of expected scores for each object of measurement can be calculated as follows:

$$
\sigma^2(aBC) = \frac{\sigma^2(b)}{n_b} + \frac{\sigma^2(c)}{n_c} + \frac{\sigma^2(bc)}{n_b n_c} + \frac{\sigma^2(ab)}{n_b} + \frac{\sigma^2(ac)}{n_c} + \frac{\sigma^2(abc)}{n_b n_c}
$$

$$
X_{aBC} = \bar{X}_{aBC} \pm z_{\alpha/2} \cdot \sqrt{\sigma^2(aBC)}
$$

where $n_b$ number of levels for facet $b$,
$n_c$ is the number of levels for facet $c$, $\bar{X}_{aBC}$ is the expected score of object of measurement a with respect to facets $b$ and $c$, $z_{\alpha/2}$ s the z-score for the desired confidence level.

\section{Conclusion}
\texttt{GeneralizIT} is a Python package that simplifies the application of Generalizability Theory in research and practice. By automating the calculation of variance components, generalizability and dependability coefficients, and conducting D-Studies, \texttt{GeneralizIT} empowers researchers to conduct robust reliability analyses with minimal coding effort in an analysis environment with which they are familiar. Currently, no Python package exists for conducting G-Theory analyses, creating a significant barrier to its widespread adoption. This work addresses this critical need by providing a user-friendly platform for balanced crossed and nested designs. While other G-Theory software solutions allow for analyses that take into account missing data and/or work with unbalanced designs, they are more user intensive and/or proprietary. We reserve these features for future work. Presently, \texttt{GeneralizIT} offers an accessible, computationally efficient solution for improving measurement reliability in diverse research environments. By democratizing the use of G-Theory, \texttt{GeneralizIT} provides a valuable resource for researchers seeking to enhance the validity and reliability of their measurement instruments and make sound statistically informed decisions.

\section{Acknowledgments}
The authors would like to acknowledge Professor Theresa Kline for her support and insights into the methodology of Generalizability theory as they were invaluable to this work. 

%Bibliography
\bibliographystyle{unsrt}  
\bibliography{references}

\begin{thebibliography}{1}

\bibitem{brennan:2010}
Robert~L. Brennan.
\newblock Generalizability {Theory} and {Classical} {Test} {Theory}.
\newblock {\em Applied Measurement in Education}, 24(1):1--21, December 2010.

\bibitem{teker:2015}
Gulsen~Tasdelen Teker, Nese Guler, and Gulden~Kaya Uyanik.
\newblock Comparing the effectiveness of spss and edug using different designs for generalizability theory.
\newblock {\em Educational Sciences: Theory and Practice}, 15(3):635--645, 2015.

\bibitem{bloch:2012}
Ralph Bloch and Geoffrey Norman.
\newblock Generalizability theory for the perplexed: a practical introduction and guide: Amee guide no. 68.
\newblock {\em Medical teacher}, 34(11):960--992, 2012.

\bibitem{bloch:2024}
Ralph Bloch.
\newblock G-string-mv.

\bibitem{moore:2016}
Christopher~T. Moore.
\newblock gtheory: {Apply} {Generalizability} {Theory} with {R}, October 2016.

\bibitem{briesch:2014}
Amy~M. Briesch, Hariharan Swaminathan, Megan Welsh, and Sandra~M. Chafouleas.
\newblock Generalizability theory: {A} practical guide to study design, implementation, and interpretation.
\newblock {\em Journal of School Psychology}, 52(1):13--35, February 2014.

\bibitem{brennan:2001}
Robert~L. Brennan.
\newblock {\em Generalizability {Theory}}.
\newblock Springer New York, New York, NY, 2001.

\bibitem{cardinet:1976}
Jean Cardinet, Yvan Tourneur, and Linda Allal.
\newblock The symmetry of generalizability theory: Applications to educational measurement.
\newblock {\em Journal of educational measurement}, pages 119--135, 1976.

\end{thebibliography}

\end{document}